# Diabolus ex Machina: Complexity, Nomadic Resistance, and the Machinery of Forced Migration in Kazakhstan

Christopher A. Hartwell

Department of International Business, ZHAW School of Management and Law, SWITZERLAND

&

Department of Strategy and International Business, Kozminski University, POLAND

chartwell@kozminski.edu.pl; harw@zhaw.ch

ABSTRACT

Organizations may perceive disorder in their environment as a threat which can impede the attainment of desired goals. In such a situation, organizations may become single-minded, obsessed, and rely on brute force to attempt to bend the external environment to their will. These efforts may be enabled by perceived power imbalances between the organization and other actors in the environment, but they may also be stymied by the complexity of other organizations in the institutional field. This paper examines such an example in Soviet Kazakhstan from 1929 to 1933, when the All-Union Communist Party (Bolsheviks) pursued a policy of sedentarization of the nomadic people. Forcibly migrating nomads to collective farms under a brutal policy of repression, the Soviet bureaucracy at all levels saw no policy as off-limits to achieving its goals but continued to run up against resistance from the complex organizational forms inherent in Kazakh nomadism. Using archival documents, testimonies from Kazakhs, and arrest records from this time period, I show just exactly how the Soviet apparatus pulled the "devil from the machine" and how, ultimately, the machines were doomed to failure. However, the human cost of forced migration continues to linger.

*Disclosures.* None.

There are no funders to report for this submission.

ACKNOWLEDGEMENTS: The author wishes to thank Sophie Alkhaled, Mariola Ciszewska-Mlinarič, Boris Vinogradov, and multiple participants and fellow presenters at the Association for the Study of Nationalities (ASN) conference in 2026 for their comments and suggestions.

**Diabolus ex Machina: Complexity, Nomadic Resistance, and the Machinery of Forced Migration in Kazakhstan**

ABSTRACT

Organizations may perceive disorder in their environment as a threat which can impede the attainment of desired goals. In such a situation, organizations may become single-minded, obsessed, and rely on brute force to attempt to bend the external environment to their will. These efforts may be enabled by perceived power imbalances between the organization and other actors in the environment, but they may also be stymied by the complexity of other organizations in the institutional field. This paper examines such an example in Soviet Kazakhstan from 1929 to 1933, when the All-Union Communist Party (Bolsheviks) pursued a policy of sedentarization of the nomadic people. Forcibly migrating nomads to collective farms under a brutal policy of repression, the Soviet bureaucracy at all levels saw no policy as off-limits to achieving its goals but continued to run up against resistance from the complex organizational forms inherent in Kazakh nomadism. Using archival documents, testimonies from Kazakhs, and arrest records from this time period, I show just exactly how the Soviet apparatus pulled the "devil from the machine" and how, ultimately, the machines were doomed to failure. However, the human cost of forced migration continues to linger.

## Introduction

Organizational change, especially when considered at the institutional level, is famous for its difficulty in executing successfully (Burnes & Jackson, 2011). While the organizational studies literature has focused on organizational change from the point of view of firm resources, leadership, and procedures (Smith, 2002; Mellahi & Wilkinson, 2004), the reality remains that organizations do not operate in a vacuum and may find that their options for change are limited by a complex and disordered external environment (Laughlin, 1991; Haveman, 1992).

In extreme situations, the external environment may be perceived as so threatening to core organizational or institutional goals as set by founders (Ormrod et al., 2007), leaders (Brunsson et al., 1996; Herath et al., 2025) or persistent legacies or organizational norms (Foroughi et al., 2025) that an organization attempts to tame this perceived disorder directly. In this situation, the objective of the organization may shift from an attempt to adapt or accommodate to expending resources to try and overthrow the external forces threatening the organization's operations. Such an approach, an extension of Lawrence and Suddaby's (2006) idea of "institutional work" - where an organization may attempt to create, maintain, or disrupt institutions in its ecosystem – may have a firm generating new internal forms and processes to resist the world outside of the organization and to shape it in a way more amenable to the organization itself (Nenonen & Storbacka, 2020).

This Nietzschean approach ("hail to thee, my will!") may find success where external environments are characterized by weak institutions or where organizations are able to both build coalitions and make change meaningful to others in the institutional field (Moser et al., 2020). In these instances, the organization may perceive a power imbalance, a "correlation of forces" (Lider, 1980) in its favor, which will enable it to succeed in this imposition of its desires. However, attempts by organizations to tame the external environment become less likely to succeed as the external environment becomes more complex and disordered: as noted by Quattrone & Zilber (2025: 1091), "periods of crisis and unsettlement invite even more efforts to make sense and create provisional realms of order, and – less positively – may also give rise to populism and authoritarian moves, which inevitably plant the seeds for more disorder and chaos." Put a different way, original shocks within an organization's external environment may create a desire to rectify any perceived disorder but attempts to impose order on systems which are complex and difficult to understand from the outside may only lead to more disorder. Rather than being Nietzschean, an organization may find that the effect is Newtonian, as organizations resort to more stringent measures in an attempt to "keep it all together" (Burnes, 2005) while the external environment then pushes back. This possibility may induce the organization to invest more and more resources in an attempt to overwhelm the outside world, with only organizations with extensive resources able to overcome resistance (Lawrence & Suddaby, 2006).

An example of a highly institutionalized order typified by complexity is nomadism (Rogers, 2019), an organizational form that relies on movement rather than permanent settlement in order to "avoid a wide range of hazards in the physical and social environment" (Dyson-Hudson & Dyson-Hudson, 1980:17). In reality, nomadism is an economic institution (Hartwell, 2013), with pastoral nomadism in particular designed to increase the utility of the herders who practice it; in this goal, pastoral nomadism is characterized by self-organization but also adheres to internal rules and norms in order to ensure the viability of the system. These rules may have appeared inscrutable to outsiders and especially to states which sprang up around the nomads, as nomadic rules and customs seemed to enshrine political volatility while also creating rigid social norms (Tolnai et al., 2020). But their utility was apparent from inside the tribe, as they elevated expertise with regard to herding and animal husbandry (Honeychurch, 2014) and recognized that “animals can make the difference as participants in collective action, which may have ramifications for the kinds of governance possible” (Kanne et al., 2024:2). What appeared to be chaotic when viewed from the point of view of outsiders actually was a well-ordered anarchy (Erasmus, 1981).

However, although nomads created unique and contextual forms of organizing, their patterns and specific organizations were heavily influenced by external factors such as the weather, the availability of pastureland, and, in more modern times, the encroachment of various sedentary states on nomadic practices. In fact, the development of the modern nation-state has been one of the greatest threats to the organization of nomadism, with countries offering policy incentives to sedentarize the population in places as far-flung as Benin, Tibet, and Central Asia (Wang et al., 2022), In extreme cases, when policy inducements were not sufficient, states resorted to direct controls and brutal forms of coercion to stop nomads from roaming, including using the state’s monopoly of violence to forcibly migrate nomads (Tapper, 1972; Casimir, 1992) - while “forced migration” is often taken to mean the displacement of persons from one territory to another, in the context of nomadism it also can mean the restriction of normal migration routes, forcing nomads into pre-determined areas and then closing off their ability to exit (MacKay et al., 2014). In this way, states were able to make order of the apparent chaos of regular migration by forcing nomads to migrate to one territory and then locking the door behind them.

This paper tells the tale of what happens when one organization – the state – confronts another complex organizational form – nomadism – and attempts to impose its will on an essentially anarchic (but not chaotic) process. Specifically, I examine the moves of the All-Union Communist Party of the USSR (the Bolsheviks) in the 1930s to force sedentarization on traditionally nomadic people in Kazakhstan. Russia's history in Central Asia *vis a vis* the nomads of Kazakhstan was long and bloody, starting with Tsarist Russia in the 18th and 19th centuries, imposing policies of Russification on the native people and which sought to overwhelm the Kazakh majority by settling Slavs on Kazakh land (Hartwell, 2023). This repression was drastically accelerated under the Soviet Union, which imposed the communist state on Kazakhs and other people in Central Asia through a so-called sedentarization drive. The goals of the sedentarization drive were manifold, first to forestall any threats to the Bolshevik regime from anything outside of communist control (Ibadildin, 2024), but, more importantly, to ensure the "success" of agricultural collectivization in line with Leninist ideals. In other words, the organizational response of the Soviet state to the many problems of fulfilling its grandiose desires in economic planning was to attempt to erase a specific institutional arrangement which existed outside of the state, and which appeared to be a threat to collectivization.

Using archival documents from within the Soviet apparatus at all levels, both in Moscow and in administrative centers in the Kazakh Autonomous Socialist Soviet Republic (ASSR), I show how the Communist Party organized its resources to forcibly migrate nomads to sedentarism no matter what the cost in human life. In particular, the reality that a cultural shift of epic proportions was attempted administratively (Saktaganova & Sakabay, 2025) resulted in a situation that necessitated Soviet officials throwing more and more resources in an attempt to conquer nomadism; what the documents reveal is that officials were aware that the monomaniacal focus on sedentarization led to countless deaths, but this problem was ignored or even welcomed to accelerate the perceived need for communist “solutions.” The imposition of a sedentary lifestyle broke traditional nomadic economic institutions and led to famine, starvation, and the persecution of millions, even as collectivization as an economic form or a political experiment faltered and failed to improve agricultural efficiency within the confines of the collective farm. Interestingly, the organizational inertia in pursuit of sedentarization extended to continued persecution and arrests of several thousands of people simply for the crime of “nomadism,” thus making the original objective (agricultural collectivization) subordinate to the organizational objective of eliminating an entire organizational form. In this sense, the archives of the Soviet Union offer a glimpse into the “diabolus ex machina” (the “devil from the machine”), an organizational choice “devised to remove the human obstacles from the path of progress” (Hollander, 1980:501).[1]

The main contribution of this paper is to theorize organizational responses to perceived disorder from nomadism in particular, and illustrate how attempts to impose order without an understanding of the “disorder” being superseded can lead to human disasters, a vindication of the supposition of Quattrone & Zilber (2025). In particular, I highlight how the machinery of evil – that is, forced migration, deportations, and their opposite number, sedentarization – was conceived, implemented, and ultimately abandoned in the Soviet Union in the 1930s in the context of nomadism. This paper is thus the first to approach the Soviet atrocities in Central Asia as a question of organizing as much as of ideology, an approach which has been used successfully in exploring other evil regimes such as Nazi Germany (Alford, 1990; Martí & Fernández, 2013) or in the Soviet Union itself, but limited to the Stalinist purges (Bloxham, 2008). Additionally, this paper contributes to the literature on nomadism by positing nomadism as its own complex organizational structure, building on the insights of Rogers (2019) and applying it in the Kazakh context. Finally, this paper also adds to the literature on forced migrations by showing how restricting nomadism functioned as a forced migration in space: much like James Scott's (2009) history of anarchy in upland East Asia, nomads in Kazakhstan had a different perception of space than the state, which exists in delineated boundaries, and this freedom of movement was inimical to the place-based policies of central planners. This paper shows how the Soviets attempted to create provisional realms of order to realize their plans, expending resources and running roughshod over nomads to impose order by forcibly relocating them to become grounded in space.

[1] Hollander was referring to Solzhenitsyn and the administration of the Gulag, another form of organizing undertaken by the Soviets to attempt to bend reality to their will, no matter the human cost. In the context of the sedentarization program, however, it still applies.

**Theoretical Development: From Disorder to the Machines of Evil**

Organizations are never an island unto themselves, as they operate within a country's institutional matrix and are influenced by and influence institutions, cultures, and procedures which wash across borders (Hartmann et al., 2022). This external environment can, at times, shift to the detriment of the organization, creating a perception (real or imagined) of threat; as Jackson and Dutton (1988) note in their seminal work, these shifts are realized as actual threats if they can impact the organization negatively, especially if they have the ability to obstruct the attainment of the goals of the organization and/or lead to monetary loss (Kovoor-Misra, 2009) or to challenge a firm's identity or reputation (Elsbach, 2003). In extreme situations, an exogenous shift in an organization's environment can even threaten an organization's prospects for survival (Probst & Raisch, 2005).

The strategic management literature has spent four decades exploring firm responses to threats, assessing the responses that organizations may take based on assessments of the immediacy and size of the threat (Osiyevskyy & Dewald, 2018). For the most part, the literature has concentrated on one of two courses of action for firms; in the first, organizations focus on internal adaptation to change, including organizational change to accommodate external shocks (Hatum et al., 2010; Hällgren et al., 2018), while in the second, firms exhibit "threat rigidity," where top management concentrates its efforts in familiar areas where it feels it actually can exercise control rather than confront the void where it is helpless (Chattopadhyay et al., 2001; Mazzei et al., 2025).

A common implicit thread in this literature, highlighted and made explicit in recent papers (e.g., Giovannoni & Quattrone, 2025), is the perception that external threats to the organization - especially if large in scale or occurring on various fronts - are out of the control of management or leaders (Mithani & Kocoglu, 2020). However, in complex adaptive systems such as a country's institutional matrix, some organizations may be more powerful than others and thus have a higher likelihood to act as change agents (Child & Rodrigues, 2011). In organizations that perceive themselves as powerful and/or having plentiful resources, such as organs of the state, the response to external threats may be very different than smaller or more "peripheral" organizations. Indeed, powerful organizations may perceive external threats as disorder and thus may attempt to change the actions of other organizations within the field by sheer force of will. Although Berger and Luckmann (1966:103) may have asserted that all organizations are an attempt to keep "chaos at bay," where organizations feel themselves to be more powerful than others, they may perceive the workings of other organizations as chaos and attempt to impose their will.

As Quattrone and Zilber (2025) note, such an attempt to impose a particular order on a particular chaos may result in authoritarianism, but I go farther to assert that such a belief in organizational dominance – or at least the perception of organizational dominance - can lead to the institutionalization of evil in the pursuit of organizational goals. There is a long and fascinating research stream examining evil and how it is operationalized via various organizational forms, originating from social psychologists who grappled with the conditions which allow evil to arise and propagate. As Alford (1990:7) succinctly put it, "it is the *interaction* between the evil within and institutions, ideologies, and historical opportunity that explains evil in the world" (emphasis in the original).

The role of organizations in propagating or transmitting evil impulses was first examined by Alford (1990) and Darley (1992), who unfortunately faced a surfeit of real-life examples crying out for a theoretical framework to explain them. Alford (1990:19), for example, explained how "bureaucracies… are not mere transmission belts for evil orders, though they may be this too [but they] may also help… organizing ways of thinking and acting that separate the individual from his own rage, and so protect the individual from the disintegrative effects of his own aggression, as well as its consequences." Ashford and Anand (2003:12), building on this insight, noted that certain attributes of organizations actually lend themselves to malfeasance: the fact that organizations require specialization helps in the furtherance of evil as "specialization not only fosters a diffusion of responsibility, it makes it difficult for any individual to comprehend (and easy to deny) the 'big picture'." Similarly, Darley (1992:208), reviewing many contemporaneous books which examined this question, notes the incremental process of organizational evil, where an individual is asked to take steps along a continuum which, seen in isolation, are tiny in nature but eventually push a person into evildoing. Thus, according to these seminal works, the keys for the organization are to push a person's boundaries slowly, almost imperceptibly, allowing for rationalization after the fact, while organizing the machine in a way that does not expose any one individual to the full extent of the evil being perpetrated; in that sense, the role of police, judge, jury, and executioner are kept very separate and distinct.

This early psychology literature has produced a small boom in academic literature across disciplines which examines the organization of evil and how heinous acts can be encouraged and institutionalized via bureaucracy, laws, and other forms of organization (*inter alia*, Bloxham, 2008; Richardot, 2014; Balfour et al., 2019; Kwok, 2021; Lukina, 2023). It is important to note at this point that organizations may actually be explicitly evil in an objective sense from the outset: the "Final Solution" in Germany during the Second World War, for example, leading to the attempted extermination of the world's Jewry, is a prime example of an organization that had evil intent from its inception. In a similar vein, the Soviet gulag system can be classified as such another organizational manifestation of evil, one which was a benign organizational apparatus but that, from the outset, had the goal of eliminating classes of people and/or extracting labor akin to slavery.

However, as Alford (1990) stresses, there need not be evil inherent in an organizational setting, although, as Adams and Balfour (2015) note, the presence of leaders displaying some pathologies are almost always connected with pushing an organization towards evil. But where an evil goal is not baked into the organization, it takes another impetus, namely a "moral inversion, in which the bad becomes good" to turn an organization from benign (as with an amateur football team, which sees its rivals as enemies) to an "eruption of evil" (Adams & Balfour, 2008:884). From the inside, participants may believe that they are actually doing good, merely engaging in organizational duties, while seen from the outside, what is happening may not be readily perceived as evil (Adams et al., 2006); this masking of evil has been labeled "administrative evil" by Adams and Balfour (1998), where ordinary workers (not inherently evil) following routines, procedures, and bureaucracy can ultimately still produce great evil. Indeed, as Dillard and Ruchala (2005) note, the instrumental rationality of an organization, i.e., the justification of the ends over the means, can make the question of "following orders" one of passive complicity rather than active participation. Much as evil works on the individual by slightly pushing away morality in favor of instrumentalism, the same effect can then happen in an organization, as "there is no one transformative moment to which we can point, but rather an incremental shift from right to wrong that is barely perceptible on a day-to-day basis" (Jurkiewicz & Grossman, 2015:3). Once the production of evil as a goal is established within an organization, where the organizational culture has changed, the reproduction of evil is institutionalized via the creation of new procedures and standard operating procedures in furtherance of the morally inverted goals of the organization (such as those that govern the Mafia, see Anderson [1965]). In this sense, as Bakan (2004) notes, evil may actually be easier in an organization because of its lack of a multifaceted nature (unlike individuals), making it exhibit a psychopathology of singular self-focus.

And one of the aspects of this focus is, once again, the organization's relation to others, an obsession that Jurkiewicz and Grossman (2015) call "organizational narcissism," namely the organization perceiving itself as the center of attention and needing to be the dominant force in society. Such a worldview means that the organization, having institutionalized evil internally, may now be driven to spread this evil in a bid to make the world a "better place." However, the sequencing need not be, in this case, from evil -> obsessive attempt to remake the world, it can just as easily run the other direction, working from the surety of the organization that its goals are correct and right, causing its culture to fan out to encompass all manner of means to justify its perceived end. With this sequencing, the existence of other organizations which may threaten the "correct" organization is problematic, meaning that these other organizations must be repressed, broken up, or destroyed (Leung, 2002). This takes us full circle to the idea of threat perception in the external environment with the caveat that the evil organization may not only perceive the threat or perceive that it is the dominant organization, but its evil nature also means that it has the will to attempt to eradicate other organizations (Bakan, 2004).

To summarize, organizations may perceive external threats as existential and, if they are powerful enough, seek to bend external circumstances to their own will rather than look inward. However, such attempts to impose order on perceived disorder can result in the unwitting (although not always) organization of evil – especially in organizations which were already trending it that direction - normalizing heinous behavior in single-minded pursuit of the organization's goals.

**Case Selection and Methodology**

In order to explore the creation of the machines of evil when faced with a complex, potentially disordered threat, this paper follows the "historic turn" in management and organization theory (Clark & Rowlinson 2004) and relies on a multi-year historical episode involving the construction of the machines of evil for forced migration: the Soviet sedentarization drive against the Kazakh nomads in the Kazakh ASSR in the late 1920s and 1930s.

Although the Soviet leadership – either in Moscow or locally - had given no warning about a desire to eradicate nomadism prior to 1928, in April 1929, as part of the drive to collectivize all agriculture throughout the Soviet Union, this attitude changed radically. In particular, a resolution was passed at the 7th Congress of Soviets of Kazakhstan that set a goal of collectivizing and sedentarizing all Kazakh nomads by 1934 to make the republic a source of food for Russian cities and the army (Piancola, 2019). Forcing Kazakh pastoral nomads to abandon their traditional routes and remain within the borders of the Kazakh ASSR and, in particular, settle on collective farms, the policy lead to widespread deprivation, slaughter of livestock for survival, expropriation of what remained, armed rebellion, and a famine which killed over 1.5 million Kazakhs from 1930 to 1933 (Pianciola, 2001; Cameron, 2016; Hartwell, 2023). At no point during this project did any of the organs of the Soviet state relent and, according to Ohayon (2013b, Interpreting and Describing the Facts section), "the famine was the outcome of a political project of brutal transformation that paid little attention to its human cost."

In order to explore the organizing around this transformation, and why Soviet policies persisted in the fate of such disastrous consequences, we are advantaged by a peculiar characteristic of most stripes of totalitarianism, but especially those concerned with central planning: they feed off of statistics, with the entire basis of the planned economy resting on input and output tables (although, as was well known, the output data could not necessarily be reliable, see Ebeling [2021]). Moreover, given the various layers and levels of the Communist Party in the Soviet Union, we also can find documentation across the various strata of the machine, with multiple reports and interpretations from Moscow to Almaty and out into the regions on the same events. Indeed, whereas one of the main issues with archival research is "silences," or biases against what is included and what is not (Decker, 2013; Bruns, 2024), much of the machinery of evil in the USSR survives in excruciating and well-documented detail – that is, even while the data collected was done specifically for practical and immediate reasons (Decker, 2025), the fact that it was dedicated to a much larger transformation project means that it kept an eye on its utility in documenting history. Given this proliferation of documentation, the short-lived thaw which threw open archival sources after the fall of the Soviet Union in late-1991 provided a treasure trove of information regarding the organizing of these machines during the communist period.

Taking advantage of this unprecedented recording of a human tragedy, Kazakh researchers and archivists affiliated with universities and the National Library and the National Assembly spent much of the 1990s and early 2000s collecting these primary documents into several compilation volumes published in Russian and, later, Kazakh (Zhandabekova, 1998; Aubakirov et al., 2005; Omarbekov, 2011; Shepel et al., 2012; Abdygaliev, 2021), with even a contribution to this historiography coming from the archives of the Federal Security Bureau (FSB), the successor to the Soviet OGPU/KGB secret police (Berelovich & Danilov, 2005). These volumes encompass a broad range of documents, including decisions and minutes from Communist Party meetings, directives issued as a result of these meetings, directives issued independently by political authorities, proclamations published in local newspapers which were intended to have the force of law, correspondence between local authorities and Moscow or across regions of the Kazakh SSR, and internal reports written by functionaries and bureaucrats detailing the "successes" of the sedentarization drive.

In addition to these primary documents, there also are voluminous arrest records from the OGPU which have been compiled by several organizations throughout Russia and Kazakhstan. In the first instance, *Memorial International*, a nonprofit society which was created in the late years of the Soviet Union to document the crimes of communism and in particular under Josef Stalin, has created an online database, "Жертвы политического террора в СССР [Victims of Political Terror in the USSR]" (https://base.memo.ru/) with the ability to search arrest records by name, occupation, location, and other relevant attributes.[2] Similarly, in Kazakhstan the State Commission for the Complete Rehabilitation of Victims of Political Repression was constituted in 2020 and has a searchable online database for prosecutions carried out under the Soviet Union; as of this writing (March 2026), only about 120,000 records of a projected 2.4 million had been digitized but they still offer a glimpse of the repression that took place under the USSR. Many of these records survive precisely due to the innocence of the accused, in that these records were collected and exhumed as part of the Republic of Kazakhstan's effort to rehabilitate "criminals" from the Soviet era. Finally, there is an "Open List" (https://ru.openlist.wiki/) which allows users to search by numerous attributes across 3,293,384 records of political repression across the entire Soviet Union. Each database has its pros and cons in terms of coverage (for example, Open List appears to only have those sentenced to labor camps, while the State Commission and Memorial databases also contain those who were executed) and user friendliness, but the arrest records provide a human face, a microhistory (Hargadon & Wadhwani, 2023) to those crushed under the machines of evil.

[2] A winner of the Nobel Peace Prize in 2022, Memorial had been liquidated at the end of 2021 by the Putin regime in Russia for "violations" of the notorious "foreign agent" law; state prosecutor Alexei Zhafyarov said, in advance of the pre-determined verdict before the Russian Supreme Court, that "Memorial creates a false image of the Soviet Union as a terrorist state. It makes us repent for the Soviet past, instead of remembering glorious history… probably because someone is paying for it" (quoted in Ivanova [2021, December 28]).

This paper uses these multi-source archival documents to tell the story of building these machines from the point of view of the builders themselves, providing an analysis of the contextualized explanation (Buchnea, 2023) of the institutional mechanisms which enabled the forced migrations while simultaneously highlighting the agency of the bureaucrats involved (Lawrence et al., 2009). In particular, we will heed the approach of Buchnea (2023) in constructing the case, limiting our analysis not only to the perpetrators but also to the years in which the forced migration took place (1929-1933) and to the geographic area of the Kazakh ASSR (with only slight forays into China and the Russian Soviet Federative Socialist Republic [RSFSR] in the context of migration that was attempted to escape the machines). Through the lens of a historical organizational study (Decker et al., 2021), we will piece together a detailed (and hopefully somewhat complete – see Buchnea [2023]) case of how forced migration was institutionalized in this place and in this time.

**Nomadism in Central Asia**

Nomadism in Kazakhstan has a long history dating back thousands of years across the Eurasian steppe, ebbing and flowing as political fortunes (such as the destruction of settlements by the Mongols) changed in the region but persistent no matter what was occurring within political institutions (Ibadildin, 2024). The type of nomadism that was most prevalent in Kazakhstan, along with the areas of Turkmen tribes and to a far limited extent, in today's Uzbekistan, is known as pastoral nomadism, where the nomads were "stockbreeders on the steppes, desert and semi-desert areas or mountain pastures, which they combined in one way or another with arable farming" (Zhdanko, 1966:601). In the Kazakh case, their variant of pastoral nomadism relied to a much greater extent than in other nomadic organization on horse riding, first for breeding and, as cattle became more lucrative a source for breeding (Salauat, 2025), as a way to achieve greater mobility when searching for suitable pastures for livestock; this reliance on horses then also allowed for the use of the horse as sustenance when it outlived its usefulness as a means of conveyance, another distinction between Kazakhs and other nomadic tribes (Chang, 2015).

Like other forms of nomadism, the social structures of nomads in Kazakhstan formed a complex adaptive system (Rogers, 2019), apparently disordered and inscrutable to the outside viewer but in reality, comprised of internal rules and norms that evolved over time and in response to life on the steppe (Hudson, 1938). While these internal orderings may have shared some similarities to other nomadic tribes in Central Asia and elsewhere, their precise form in Kazakhstan was unique, as "pastoralism involves contingent responses to a wide range of variables in the physical and social environment" (Dyson-Hudson & Dyson-Hudson, 1980:17). In particular, the Kazakhs developed a specific set of social and economic institutions to support nomadism in the Kazakh lands, creating a multitiered framework based on family ties and lineage (Hartwell, 2023).

The overarching system of nomadism was built on the *zhuz* (horde), three separate tribes based loosely on geographic regions, further stratified into "white" and "black" bone, depending on the lineage of the tribe itself (Hartwell, 2023; Sailaubay & Zhanbossinova, 2024). However, at a more local level, under the aegis of a specific horde, Kazakhs were organized into tribes, clans, and then extended families, the *aul* (Guirkinger & Aldashev, 2016); of these, the clan and in particular the *aul* were most important, as the *aul* consisted of anywhere from 5 to 80 yurts but which mainly consisted of between 10 and 30 yurts at one time (Ohayon, 2004) and was instrumental in assigning property rights for winter pastures (Guirkinger & Aldahev, 2016). Within the *aul,* the nomadic leader and respected elder was known as a *bai*, generally wealthier (in terms of livestock) and qualified along the lines of metrics chosen by the particular *aul* (Sailaubay & Zhanbossinova, 2024). This role of the *bai*, as a local leader, was strengthened during Russian colonialism in the 19th century, as the move towards livestock meant that some farming was necessary to create supplementary feed (Kerven et al., 2021), creating pockets of semi-nomadism that Russians used to co-opt the *bai* for their own administration. At the same time, the reliance on wiser, older nomads was necessary as Russian settlements in the north cut off traditional migratory routes, making the Kazakh social network more reliant on more local authority (Ibid.).

Despite the encroachment of Russian imperialism and the move towards cattle (which were less likely to sustain long-range migrations), nomads in Kazakhstan still followed traditional seasonal patterns, moving from the steppes in the summer to winter pastures sometimes more than 200 km away in the winter (Guirkinger & Aldahev, 2016). Ingold (1986:183) described how these migrations may have looked to outsiders, noting that "[nomadic] movements can appear quite irregular and erratic, but in arid regions may in fact be closely attuned to incidence of rainfall." In fact, the logic of Kazakh nomadism was reflected in the language that the Kazakhs used, as "in Kazakh, the verb *kôšu* (to nomadize) implies stable movements toward known places, following customary routes. If the movement deviates from the usual routes, in case of danger for example, the verb *auu* is used, which connotes an idea of disorder, deviation from a norm, and imbalance" (Ferret, 2014:959).[3] Nomadism in the Kazakh context was thus predicated on a stable patrilineal social order that had seasonal migration based on local knowledge and experience, following climatic, rather than political, boundaries.

[3] Translated from the original French: "En kazakh, le verbe kôsu (nomadiser) implique des mouvements stables vers des lieux connus, selon des trajectoires coutumières. Si le déplacement quitte les trajectoires habituelles, en cas de danger par exemple, c'est le verbe auu qui est employé, lequel connote une idée de désordre, d'écart par rapport à une norme et de déséquilibre."

**The Threat to Bolshevism**

Much has been written on the organization of the Communist Party of the Soviet Union and how it operated in practice across various strata (Daniels, 1957; Getty, 1987; Schapiro, 1987; Fitzpatrick, 1988) , and thus this section will not attempt to rehash this voluminous literature but merely illustrate the key tenets of Bolshevism in the Soviet Union as can be seen in the policy towards the nomads. And indeed, it is fairly easy to see why nomadism in Kazakhstan, as elsewhere, was perceived as a threat by the authorities, as (for the Soviets) it was a way of organizing which ignored political borders and thus threatened the authority of Moscow.

This authority, of the Bolsheviks (formally, the All-Union Communist Party as of 1925), was structured in a much different pattern than the Kazakhs; named a "mono-organizational society" by Rigby (1999), the centrality of the Communist Party in the Soviet Union created a fountain from which all decisions flowed down to various hierarchies, a system based in theory on Bolshevik leader Vladimir Lenin's so-called "democratic centralism" but which was far more "centralized" than "democratic." While the Kazakh *aul* elevated the *bai* on the basis of lineage and experience, the Bolsheviks created a two-track administration which put administrative and technical bureaucrats alongside non-technical party functionaries designed to mobilize the masses (von Beyme, 1975). These functionaries, the so-called *apparatchiki*, were elevated not because of their expertise but because of their ideological fervor, steeped in the ideological strictures of communism as interpreted first by Lenin and then by Stalin. As Hill (1986:26) described it, "apparatchiki possess enormous power, which has accrued over the decades… [as] those who claimed understanding of the needs of society, an understanding supposedly based on a 'special relationship' with the ideology, placed themselves in a position of unchallengeability [*sic*] as far as the masses were concerned." Three decades earlier, Daniels (1953:159-160) concurred that leadership in the Soviet Union was "dependent for its security on the Party's organizational machinery; the latter was thus becoming the real locus of power" (Daniels, 1953:159-160).

The elevation of communism as the state religion, the prohibition on "factionalism" and dissent, and the enforced infallibility of the decision-makers in the bureaucracy meant that the Soviet system had created machines for enacting the wishes of the top leadership without any inclusion of the masses being affected. More importantly for our purposes, the bureaucratic apparatus which was created also was very sensitive to threats both from within and from outside the Party and had additional mechanisms to scan the horizon for threats to the regime. Internally, there were a number of verification campaigns to vet Party members (Fitzpatrick, 1979), as well as a Central Control Commission which conducted purges within the Party for "factionalism" and other sins (Getty & Naumov, 1999). Far more nefarious to the communist dreams of a new society, however, were external threats, that is threats within Soviet society but outside of the Party apparatus; this included rich landowners (*kulaks*), counterrevolutionaries, Whites (defeated supporters of the restoration of the monarchy), and assorted saboteurs and "wreckers" of the new Soviet state (Harris, 2016). To combat these threats (real and perceived), the Soviets relied on an extensive security apparatus, centered on the Joint State Political Directorate (OGPU, standing for *Объединённое государственное политическое управление*). The OGPU was the secret police arm of the Soviet state and the successor to the State Political Directorate (GPU), as well as the precursor of the perhaps better-known People's Commissariat for Internal Affairs (*NKVD, Народный комиссариат внутренних дел*).

To ensure the safety of Party and the country, the OGPU relied on orders from the Central Committee for its direction and, as political scientist James Q. Wilson (2019) noted in the context of other bureaucracies, its own internal standard operating procedures (SOPs) for implementation. From the outset of the Bolshevik revolution, the security forces had been used in a series of repressions directed by the Soviet leadership, including the "Red Terror" after the Civil War, a series of show trials in the 1920s, and then the "dekulakization," collectivization, and sedentarization programs starting in 1928. As Shearer & Kaustov (2015) note, from 1928 onward, whereas the OGPU was previously focused on industrialization in the cities, the failure of communism in the countryside meant that their attention was now being focused on rural areas of the Soviet Union, and their funding and power increased as they were expected to break the backs of the peasants.

In this reorientation, the OGPU had a set of tried-and-true SOPs, including arrests, mass purges, torture, extra-judicial proceedings, deportations, and summary executions, all in pursuit of the Party's objectives. And during times when the Central Committee had decided that repressions needed to be stepped up – as in the programs running through 1929 and into the 1930s - the OGPU was given quotas of arrests that they had to achieve; in these instances, they relied on simplified procedures in order to enhance their efficiency, including the substitution of confessions for evidence (Gregory,1999). In any event, the role of the OGPU was clearly stated by Feliks Dzerzhinsky, the notorious founder of the Cheka (the first iteration of the security forces), in that the security forces were meant to be the "fighting arm of the Party," using their SOPs in achieving the goals of the Central Committee (Leggert, 1981). In practice, this meant "altering social and economic relations through administrative violence" (Shearer & Kaustov, 2015:90), or, as Shearer (2001:505) explicitly stated, “solving problems of mass social disorder became synonymous with the political protection of the state and defined a major priority for political leaders and high officials of the OGPU.” Thus, any threats to the Party required “mass social re-organization… [with] the functions of social order and state security… linked” (Ibid.).

**Forced Migration in Kazakhstan: the Soviet Drive for Sedentarization**

As noted above, the sedentarization drive in the Kazakh ASSR was not originally a piece of Soviet policy towards nomads but did logically follow on from Tsarist imperial policies in Central Asia. In the second half of the 18th century and into the 19th century, as Russian imperialism pushed deeper into the steppes on Russia’s southern border, Russia attempted to pacify the nomads, elevating weaker rulers and, in a candid statement from Tsarina Catherine, explicitly noting in a letter that Russian policy was the “desire to suppress their [Kazakh] self-administration” (re-produced in Malikov [2019:49]). While nomadism was certainly inconvenient to the Russian Empire, it was not subjected a frontal assault for eradication; instead, the Tsarist policy was more focused on supplanting Kazakh legal traditions with the Russian legal code, relocating Russian and other Slavic settlers to the “virgin lands” of the Kazakhs, and a multi-tiered policy of “Russification” which eliminated both cultural aspects of the Kazakhs and their language (Hartwell, 2023). It was assumed, as part of this assimilation into the Russian Empire, that “the internal support structures of nomadism had been destroyed and an irreversible process of economic change had begun which could only end in the sedentarization of the Kazakh nomads” (Olcott, 1981:20).

While the policies of Russification, the bloody and ultimately suppressed revolt in 1916, and the success of the Bolsheviks in the Russian civil war led to the Kazakhs being re-absorbed into the communist Union of Soviet Socialist Republics (USSR), it was not until the late 1920s and the ascent of Josef Stalin as General Secretary of the Communist Party that nomadism – stubbornly clinging on and not withering away at the speed that the Bolsheviks would have liked - was explicitly targeted. The opening salvo against nomadism was launched by People's Commissar of Agriculture K. Toktabayev, during the 7th Congress of Soviets of Kazakhstan in April 1929, who announced that "the first task in expanding the sown area, to which we must now turn our most central attention, is the issue of the settling of the Kazakh population." This declaration was the basis for the resolution opening the sedentarization program (VII All-Kazakh Congress of the Soviets, 1929), which stated quite clearly that "the development of grain farming in the region rests primarily on the problem of the settling of the semi-nomadic and nomadic population... in all parts of the republic" (*Ibid.*, p. 247).

Like the previous Tsarist attempts to erase Kazakh culture, the sedentarization drive served multiple Soviet goals simultaneously, but two mutually reinforcing ones stood out. In the first instance, tying the Kazakhs to the land was seen as a way to increase agricultural output, a way to augment the collectivization drive to bring all agricultural production into collective farms and turn Kazakhstan into another "breadbasket" like Ukraine. Collectivization was thus the organizational form that the Soviet machine (and above all, Josef Stalin) selected for increasing productivity but it required sedentarization to be effective, especially given Soviet planners' high hopes for crop and meat production from the Kazakh ASSR; as Uraz Isaev, a prominent Kazakh Bolshevik, mentioned at the 15$^{th}$ Party Congress in Moscow, "we need to develop land in the vast expanse of Kazakhstan, populated by various nationalities. Here, we have a number of shortcomings, both in terms of our understanding of the issue from the perspective that all available land must be used as quickly as possible" (Gosizdat, 1928). Although Kazakhstan's semi-arid steppe lands were perfectly suited for animal husbandry – and Soviet plans called for Kazakhstan to "rival Chicago" in terms of its meat-producing abilities (Kindler, 2018) - there was hope that the black soil (*chernozem*) regions of northern Kazakhstan would be able to support massive amounts of grains, particularly wheat, while the southern lands (more adjacent to sources of water) could support crops necessary for supporting cattle, such as hay (Kabuldinov et al., 2023). In this way, a key tenet of the Soviet system, starting from "War Communism," could be fulfilled, namely the hub and spoke system, where agriculture was collected *en masse* from the spokes of the countryside and transported to the cities. This approach, a "tribute" from the countryside to the cities (Ellman, 1979), was seen as a way to rapidly support industrialization in urban areas but also to quash any thoughts of uprising or discontent (without paying much attention to what was happening in the countryside, see Conquest [1986]). As Pianciola (2001:239) said, these policies were really "a cover for the indiscriminate pillaging of the entire rural Kazakh population."

With the Soviet bureaucracy and planning organs oriented towards imposing their organizational model on the country, other forms of organizing were not to be tolerated. The key problem was the resistance of the Kazakh *aul,* as noted above, the key organizational unit of Kazakh nomads; according to Bratcher (2025:44), "Soviet ideology had done little to penetrate the social and economic functioning of the *aul* by the time of [Filip] Goloshchekin's arrival as Communist Party chief" in the Kazakh ASSR in 1924, and the Bolsheviks blamed this lack of success on the richer nomads, the *bai*, who had considerable power in the traditional clan structure of the *aul* (Ohayon, 2013a).[4] The goal for the Soviets was thus to break these traditional forms of organizing and, in particular, the nomadism which the *aul* (and the *bai*) enabled. Lest there be any misunderstanding on what this entailed, the First Secretary of the Kazakh Regional Committee of the All-Union Communist Party (Bolsheviks) Filip Goloshchekin reiterated two years into the sedentarization policy that "an old, backward, nomadic and semi-nomadic village is dying, must die. In its place should come, comes a new Kazakh collective farm village with settlement, European-type buildings, with a livestock farm of the type of commercial farms" (Goloshchekin, 1931). The policies were predicated precisely on achieving this goal, with "the procurement quotas assigned to individual administrative regions were divided in such a way that it was principally the nomads and semi-nomads who bore the brunt of the provisions" (Pianciola, 2004:159). Just as Solzhenitsyn would refer to the series of concentration and labor camps across the Soviet Union, the gulag, as "our sewage disposal system," the sedentarization drive would attempt to bring the wealthy *bai* under Russian control, while the collectivization process would confiscate their belongings and break them.

Unfortunately for the Soviet central planners, the program of eradicating nomadism was never going to be easy. As of 1930, approximately 540,000 out of 700,000 farms in the Kazakh ASSR were classified as "nomadic" or "semi-nomadic" (Saktaganova & Sakabay, 2025), with 38% of the farms classified as "settled" were only settled during the previous decade. As noted above, the main arm of implementation of the program, in addition to the regional and local committees, was the OGPU, which was given a free hand to supervise both sedentarization and the collectivization drive. Originally empowered in the decision for collectivization from the 15th Party Congress

[4] In some Soviet documents, *bai* is used interchangeably or hyphenated with the Soviet favorite *kulaks* to refer to well-off peasants.

(Gosizdat, 1928), as stringent measures became more "necessary," it gained progressively more powers in rural areas and especially in Kazakhstan. These powers, already fearsome in the context of Stalin's repressive rule, ranged from classifying anti-sedentarization and anti-collective measures as "counter-revolutionary" (thus allowing for use of the death penalty) to an explicit decree to utilize the OGPU's transport and labor camp apparatus to "liquidate" the *kulaks* as a class (RGASPI, 1930, January 30).

Before it came to liquidation, the main instrument in the beginning phase of the sedentarization campaign was confiscation, including the expropriation of cattle and other livestock, alongside eviction from specific tracts of land and forced migration into either designated areas or, more commonly, pre-designated collective farms. This campaign was announced in Kazakh newspapers on June 22, 1928, with the caveat that confiscation would only affect the wealthy, and thus collective farms need not fear (and should not act hastily):

> *Recently, anti-Soviet elements have been spreading provocative rumors about the nature of the confiscation of property of rich farms. Under the influence of these rumors, there have been instances of livestock being sold off by labor-intensive livestock farms, which leads to an unacceptable weakening of these farms and creates a threat to the further successful development of the national economy. The Presidium of the Kazakh Central Executive Committee informs all workers of Kazakhstan that the development of a confiscation law is being conducted with the aim of evicting and confiscating the property of those largest rich farms, which, as of April 1, 1928, had herds numbering in the thousands and were clearly hostile to Soviet power; First and foremost, this will affect the most prominent bais from among the descendants of former khans and sultans. The*

> *Presidium of the Kazakh Central Executive Committee (KazCEC) urges all executive committees and councils to immediately and widely notify workers of this, holding those spreading false and provocative rumors strictly accountable* (Soviet Steppe, 1928, June 22).

Already prior to this publication, local Soviet organs in the Kazakh ASSR had enthusiastically embraced the directive from Moscow for confiscation. For example, the Aktobe Provincial Committee of the Bolsheviks noted in April 1928 that "the economic conditions of the country now demand the maximum exertion of all forces and resources on the question of grain procurements, since 'in order to meet the minimum requirements for grain we need to procure at least 100 million poods for this quarter' (from the directive of the Political Bureau of the Central Committee of the CPSU (Bolsheviks)). 7 million poods should [thus] be procured throughout Kazakhstan" (APRK, 1928, April 29). But the goal of breaking the back of the nomadic population was always stressed in official documents and public pronouncements, as a newspaper article in late 1929 castigated the Kazakhs with the headline, "animal husbandry is still in the hands of kulaks and bai" (Soviet Steppe, 1929, December 24).

Over the next four years, the machines worked intensely to ensure a steady stream of confiscations, leaving the nomads without the tools to remain nomadic. For example, a report from the Kazakh Regional Committee from March 1930 noted that 429,776 households had been forced onto collective farms, with approximately 3,189 cows, 415 horses, and 135 oxen seized from Alma-Ata and Uralsk alone (APRK, 1930, March 20). As is the case with every policy sold to target specifically "the rich," however, the sedentarization and collectivization drives went far beyond just affecting the "wealthy;" a brave anonymous letter written to Josef Stalin himself from a nomad in the Akmola region of the Kazakh ASSR bluntly said, "they told the poor that they would give you seeds from the state – just accept the plan, give you rations, give you tractors, etc. If something did not work, then they began to threaten with reprisals. With a heavy sigh, the peasants took it. A nightmare began: trials, confiscation, arrests" (APRK, 1930, 31 May).

As a complex system oriented towards migration as a way to provide sustenance, the Soviet attempts to eradicate nomadism naturally met with resistance, as Kazakh nomads "voted with their feet" and continued to migrate to find increasingly scarce food. In a clearest sign of nomadic attempts to escape the sedentarization project, approximately seven thousand nomads attempted to enter China over the entire course of sedentarization, with approximately 1,478 people (and 23,214 cattle) stopped by border guards in 1929 alone (and with nearly every head of cattle seized by the state).[5] Later data published by the Kazakhstan Regional Committee of the Bolsheviks showed that migration to China continued but were supplanted by more "local" migration, with a total of 3,436 households and 187,909 cattle migrating to Uzbekistan after a local census in 1930 and an additional 7,371 households (and 319,725 cattle) moving to Turkmenistan in-between the census and August 5, 1931 (APRK, 1931, August 9). By January 1932, the OGPU stepped up its efforts to prevent nomads from even getting to the border by increasing domestic surveillance and arrests:

> *1) It is proposed to conduct disinformation campaigns among the nomads in order to prevent flight to China, the return, and the extradition of the leaders of the migrations, and*

[5] Calculation of cattle and nomads stopped taken from data provided in "Сведения полномочного представительства ОГПУ по Казахстану «Об откочевках казахских хозяйств в Китай и количестве задержанных при переходе границы за 1929 г.». 1 февраля 1930 г." [Information from the OGPU Plenipotentiary Representative for Kazakhstan "On the migration of Kazakh households to China and the number of those detained while crossing the border in 1929." February 1, 1930], published in Kondrashin (2011:205-206).

*the flight itself; 2) the organizers and leaders of the nomadic movement, as well as all the bais, are identified among the nomads, material is collected on them, followed by processing and transfer of cases for consideration by the troika of the Plenipotentiary Representative of the OGPU; 3) connections with abroad are probed and the presence of agents of foreign bais counterrevolution among the nomads is established; 4) it is proposed that the district committees of those districts from which those who arrived migrated raise the question of urgently sending party workers to work on organizing the return and providing material support for the return (APRK, 1932, January 23).*

The efforts to control the border may have deterred some nomads, but as the scale of the famine worsened, more attempted to flee to other areas of the Soviet Union and again to China. In 1933, the OGPU noted in a Top Secret document, "due to food shortages in the Dzhambeitinsky and Dzhanybeksky districts, which resulted from the improper implementation of grain procurement plans based on false information provided by collective farms and the mass theft of grain by 'false positives,' in December, cases of migration from one district to another and migration to adjacent districts in other regions were recorded" (SAWKR, 1933).

Moreover, the deteriorating conditions of the nomads was catalogued by various strata of the Communist Party but with little done to correct course, mainly because the suffering of nomads was secondary (or worse) to the goal of fulfilling the plan. As an example of the dire straits which the nomads found themselves in, according to an OGPU report "on the mass arrival of Kazakhs into the territory of the Middle Volga Region" in February 1933 (Figure 1), the officers noted that "The arriving population is in an extremely exhausted condition. Among the arrivals there are a considerable number of completely destitute households, having no livestock, no property, and no means of subsistence. A large proportion consists of women, children, and elderly persons. The overwhelming majority of the arriving Kazakhs have no documents. Many arrive spontaneously, without any organized direction" (GARF, 1933, February 12). This state of affairs also hindered other plans for collectivization, as the hungry and weakened nomads saw "typhus and smallpox began to rage… in the Slavgorod district alone on March 5, 117 deaths were registered on this basis. In search of food, the nomads move deep into the Western Siberian region, carrying the infection with them" (CAFSBRF 1932, April 1).

Whereas, prior to the famine and confiscations, nomads were driven to migrate based on the needs of their herds, now their movements were for survival. According to another report from the OGPU from August 1932, "the unsettled portion of the nomads continues to subsist on carrion and garbage from garbage pits. There have been cases of nomads attacking canteens, creameries, and workers' and employees' apartments, threatening to kill them and taking bread, milk, and other food products" (CAFSBRF, 1932, August 8). Children were also not spared the privation created by the sedentarization project:

> *The sanitary and living conditions of Kazakh nomadic children are dire (Karachi village). Some of them live in dilapidated dugouts; there is no heating, it is dark and damp, and it is very crowded—there are from 11 to 17 people per 10 square meters. As for clothing, the condition is very poor; you can see half-naked people. The children are naked, barefoot, covered in rags, torn to shreds. In terms of health, they are all emaciated, some are swollen from malnutrition. The children have scabies, ringworm, and boils. Lice are rampant. There are children under 6 who haven't been vaccinated against smallpox. Child mortality is observed* (GARF 1932, April 14).

In an unprecedented and honest assessment, several reports from across the Soviet machine noted that cannibalism had also begun to be practiced by starving nomads (APRK, 1932, August 4; CAFSBRF, 1932, July 7; CSARK, 1932, December 1) and, unlike reports from the North Caucasus and Ukraine which had mentions of cannibalism removed when transmitted to higher ups, the reports of cannibalism among the nomads remained in the reports detailing their plight. Indeed, the reports often contained graphic details, such as how desperate nomads were more likely to attempt to eat those who died of smallpox rather than those who died of hunger, as the smallpox sufferers were likely to have more fat on them (CSARK, 1932, December 1).

Despite these conditions – and their being catalogued by the Soviet machines! - the Bolsheviks saw the continued attempts at migration to escape starvation not as a question of survival but once again as evidence of a counter-revolutionary conspiracy. In an internal Bolshevik memo at the end of 1931, the regional committee conceded that there may have been "major mistakes made in the practice of planning and managing the economy of the district" and "excesses [!] in the conduct of economic and political campaigns," but that one needed to look at the "bai provocation and influence based on clan ties existing between the villages and the Kazakh clans that had previously migrated to China" (APRK, 1931, December 31). A representative of the Border Guards concurred in a report written three weeks later in January 1932, noting that "the whole process of flight should be considered as a provocation of the *bais*, fanning rumors of famine, provoking isolated cases of famine, and exploiting the excesses of local workers" (APRK, 1932, January 23). A telling memo from the full-time representative of the OGPU in Kazakhstan utilized this exact language, noting that, even though detained nomads told the OGPU that "the reasons for the migrations… are the search for food due to complete material insecurity," the memo's author said that "the real causes of the nomadic movement [are]… in connection with the presence of a significant Bai stratum among the nomads [and thus] the whole process of flight should be considered as *a provocation of the bai* [emphasis mine]" (APRK, 1932. January 28). Thus, despite the nomads explicitly saying what the reason for their migration was, the entire Soviet apparatus appeared to know what was *really* behind their actions and united behind the same wording.

However, the acknowledgment in the 1931 memo that perhaps the entire sedentarization project was not going as planned was a consistent theme throughout the entire drive, one which should have given the Soviet apparatus pause but apparently did not. Already in 1930, in a letter from the Chairman of the Central Executive Committee of the Kazakh ASSR, the "excesses" of collectivization were noted, with the Chairman castigating the Secretary of the Kazakh Regional Committee that "the population is exclusively engaged in nomadic cattle breeding, the collective farms are organized administratively, despite the fact that these villages did not sow, they were subject to grain procurements and the collection of the seed fund. Moreover, the grain requisition plan and the collection of the seed fund were not only communicated to the middle peasants, but also to the poor peasants, and therefore the population fulfilled its debts by selling cattle for a price, and thus the available number of cattle was reduced to 50%, but it was not checked at all" (APRK, 1930, April 21). In October 1932, the Deputy Chairman of the Council of People's Commissars of the Russian Soviet Federative Socialist Republic, Turar Ryskulov (himself a Kazakh) wrote a report to Stalin directly which noted that livestock before the sedentarization program (in 1928) numbered approximately 32 million heads (and Kazakh authorities thought the number might be closer to 40 million), but as of February 1932, there were only 5.397 million left, a decrease of 83% (RGASPI, 1932, October 6). This was a major problem for the drive, as, although "Stalin and other members of the Central Committee had little concern for the needs of starving Kazakhs […] the republic's precipitous drop in livestock numbers, which was apparent to everyone on the Central Committee by 1932, directly affected the regime's economic interests" (Cameron, 2018:154).

Ryskulov also noted that the sedentarization program had created problems with grain procurement, as there was no recognition of the fact that much of central Kazakhstan was unsuitable for growing grains, and even where settlement had proceeded apace, yields were low and massive irrigation was required.[6] An OGPU document from November 1932 lamented that "As of October 5 of this year, up to 30% of the total sown area of grain is still unmowed. According to official data, in the haystacks and windrows along the edge, there is a harvest of 500 thousand hectares of wheat, which begins to rot and germinate. The September grain procurement plan has been fulfilled by less than 50%. They did not give the necessary increase in the rate of grain procurement in October either. As of October 5, the annual plan for the procurement of grain was fulfilled only by 30.5%" (CAFSBRF, 1932, November 5). This failure was noticed in Moscow, as Stalin and Molotov (the Chairman of the Council of People's Commissars) sent a telegram a few days later to the Bolsheviks in the Kazakh ASSR threatening the local Committee and Council of the People's Commissars with retribution:

> *Despite the warning of the Council of People's Commissars and the Central Committee, grain procurement in the republic continues to decline. References to crop yield figures as the reason for the failure to fulfill the established plan cannot be taken into account by the Central Committee and the Council of People's Commissars, since these figures are clearly*

[6] For his candor, Ryskulov was one of the hundreds of thousands murdered during Stalin's purges in 1937.

> *prevalent and calculated to deceive the state…. The Central Committee and the Council of People's Commissars warn you that if a real breakthrough in grain delivery is not organized in the republic in the shortest possible time, they will be forced to resort to measures of repression similar to those in the Northern Caucasus. In order that you may have a clear idea of these repressions, we are sending you a resolution of the Northern Regional Committee* (APRK, 1932, November 8).

Faced with the faltering of the program in returning the results promised, the problem that was identified by many layers of the Soviet machine, as many organizations and especially politicians before and since have asserted, was that the communications surrounding the policy were the problem - not the policy itself. According to the minutes of the meeting of the Bureau of the Kazakh Regional Committee of the All-Union Communist Party (Bolsheviks) in March 1930, "taking into account the possibility of reducing the number of fictitious collective farms in the course of correcting the excesses, at the same time the main emphasis of the nomadic districts should be shifted to the creation of all the economic and political prerequisites for a really healthy mass collectivization, which is inconceivable without strengthening, strengthening the simplest associations of agricultural cooperatives, measures to increase the marketability of animal husbandry, the widespread deployment of work with the poor, farm laborers, and the restructuring of the work of the Party, Soviet, and Komsomol" (APRK, 1930, March 21).

Regardless of the "excesses" of the Soviet machines in pursuing sedentarization, the Soviet apparatus was also quick to assign fault to the nomads for their own condition, in particular their inability to follow the Party's wishes for sedentarization. The OGPU noted that there was some success in repatriating Kazakhs who had fled for West Siberia in Russia, but that those who remained nomadic only had themselves to blame: "Due to efforts to return Kazakh nomads to Kazakhstan, their numbers in the West Siberian region have significantly decreased, and no further influx to the West Siberian region has been observed, with the exception of isolated cases. According to the latest data, out of 45,000 Kazakh nomads, only 13,000 remain in the West Siberian region, of whom only 1,400 have found employment. The remainder, due to the *blatant disregard of district Soviet and Party organizations and unwilling to improve their situation*, are destitute and experiencing acute food shortages" (CAFSBRF, 1932, August 8). In the same document, the OGPU basically noted that the nomads deserved what they received:

> *As a result of the aforementioned robberies, raids on canteens and apartments, and the violent seizure of food, a hostile attitude toward nomads has emerged in a number of districts, accompanied by beatings, lynchings, and murders of nomads. Particularly brutal abuse of nomads was recorded in the Kupinsky district, where leaders of grassroots Soviet organizations, state farms, and collective farms set fire to nomads' yurts, subjected them to severe beatings, torture, and other forms of torture.* (Ibid.)

The only standard operating procedure that the Soviet state was willing to try to both correct these excesses and convince nomads to give up their nomadic ways was an increase in repression, using the heavy hand of the state for those who did not submit. Throughout the Soviet Union, Article 58 of the Soviet criminal code ("counter-revolutionary activity") was notorious for its malleability and could be stretched to cover any crime that the Soviet machine felt was a threat to the system – and which then warranted much stricter punishment, including the death penalty (Berman, 1972). In a long and rambling resolution of the Politburo in February 1930 (also marked top secret), the highest levels of the Soviet leadership considered the issues with collectivization which had been seen in the non-Russian republics of the Union and advocated that richer Kazakh nomads be considered for the ultimate punishment under Article 58 if they resisted (and were to be forcibly resettled even if they did not):

> *In addition to the resolution of the Central Committee of January 30, to carry out in relation to kulaks (bais) in the areas of complete collectivization, in accordance with the characteristics of the national districts, the following measures: the counter-revolutionary Bai-kulak activists (organizers, participants, accomplices of banditry and counter-revolutionary organizations) to be imprisoned in concentration camps, without stopping at the use of capital repression (1st category) in relation to particularly malicious elements; b) the rest of the clearly kulak (Bai) element is to be resettled outside the districts of complete collectivization on the worst lands in areas remote from the border, but within the boundaries of the given national republic* (RGASPI, 1930, February 20).

In terms of its relevance to the nomads of the Kazakh ASSR, there was a definite pattern to the application of Soviet "justice," as those arrested at the border (e.g., those trying to flee to China) were often (but not always) sentenced to labor camps, while those arrested away from the border were typically shot. An example of this policy comes from the February 1930 meeting of the Semipalatinsk Bolsheviks, who noted (in top secret minutes) that defeating the migratory patterns of the nomads would require the political apparatus "to instruct the OGPU to intensify repressions against the bais and kulaks who are agitating and working for migration to China," as well as confiscating any hunting weapons that the nomads had (CDNI, 130, February 4). Similarly, the top secret resolution of the East Kazakhstan Regional Committee of the Bolsheviks in January 1933 went further in preventing nomads from using their own property, proposing that "the heads of the district departments of the OGPU to use Chekist and other measures to identify specific perpetrators of theft, unauthorized slaughter [of livestock], and a number of directors and other employees who have allowed the squandering of state property, to immediately try them in a show trial, using execution by firing squad" (RGASPI, 1933, January 29). Already by 1932, when the failures of sedentarization and collectivization were apparent, Goloshchekin urged his comrades to step on the accelerator, "not only can we not abandon the methods of force, violence, and harsh repression toward the class enemy, we must step up the efforts" (quoted in Kindler [2018:94]).

The recourse to repression was accompanied by a familiar theme in Russian history, namely that the nomadic resistance must have been directed by external enemies (or perhaps just by internal enemies of the revolution). This reading of the situation betrayed a severe lack of understanding (or perhaps willful ignorance) of the complex nature of nomadism, and in particular the spontaneous and (to the Soviets) disordered processes could not have emerged without external agents directing the simple nomads. The presence of acts of rebellion and banditry, in addition to the aforementioned voluntary migration, confirmed this suspicion that the nomads were organizing specifically against the Soviet state. As early as "the autumn of 1929 there were cases of mass movement and discontent," including a number of instances where "the masses… organized themselves into a number of gangs aimed at overthrowing the Soviet power and forming a khanate in Kazakhstan" (APRK, 1930, June 10). The OGPU also noted "the dissemination of provocative rumors of an insurrectionary nature about the imminent death of the Soviet power" (APRK, 1930, March 25) in large swathes of Kazakhstan, with even "poor peasants who sympathized with the kulaks [being] accused of anti-Soviet agitation during the repressions" (APRK, 1930, February 19). The resolve of the Soviets in extinguishing this resistance was clear, as a telegram from Central Executive Committee of the Kazakh ASSR to the General Secretary clearly stated that "the 2nd village of Chokbar district is still militant, fearing reprisals, and therefore poses the question: reprisals are inevitable, and therefore they must die with arms in hand" (APRK, 1930, April 21). This approach was put into perspective in a comprehensive document by the OGPU in late March/early April 1932 not just for the Kazakh ASSR but across the entire Soviet Union; however, the section on the Kazakh ASSR focused specifically on the nomadic population and its resistance to sedentarization, noting that multiple attacks had occurred on the slaughterhouse of the *Soyuzmiaso* (literally, "Union Meat") facility, with one assault by as many as two thousand Kazakh nomads who were desperate for food (CAFSBRF, 1932, April 1). Even three years into the sedentarization program, resistance continued, driven by the *bai* and especially poorer nomads:

> *Grain procurements in the [Kazakh ASSR] are taking place in an atmosphere of active resistance of bai-kulak and anti-Soviet elements, expressed in the organization of mass theft and concealment of grain, refusal to fulfill transportation plans, threshing and delivery of grain, damage to harvesting equipment, etc. ... In the Kzyl-Orda and Kazalinsky districts, where the percentage of fulfillment of the grain procurement plan ranges from 5 to 18%, there is already an aggravation of food difficulties. In connection with the aggravation of food difficulties, in these areas the migrations of the indigenous population to other regions intensified again: from Kazalinsky river to Central Asia, out of the remaining 6517 farms with a livestock of 13.7 thousand different cattle, in September-November of this year, 3400 farms migrated, stealing 8 thousand head of cattle, and the migrations gradually covered the other areas listed above. In the kolkhozes of the Kostanai District, the tendencies to leave the kolkhozes and leave the district intensified* (CAFSBRF, 1932, December 7).

Seeing the shadowy hand of conspiracies afoot, the OGPU appeared to be pushing on string in actually eradicating them; over a seven-month period, the OGPU claimed to have "liquidated" 10 groups in Kazakhstan who had "organized" banditry, especially relating to the slaughterhouse. However, from the point of view of the OGPU, these nomadic groups were powerful indeed, as "in all their wrecking activities [they] artificially created excesses, organized widespread theft and concealment of grain, placed Bai elements at the head of collective farms, carried out the enlargement of collective farms in the process of grain procurements. carried out the resettlement of entire villages of collective farms to remote areas and thus introduced confusion into grain procurement plans and artificially created food difficulties and hunger" (CAFSBRF, 1932, April 1). Faced with the failure of sedentarization program, the OGPU had no choice but to turn their nemesis the nomad into a superhuman being. Indeed, the paradox of the nomads is that they may have been simple and part of a dying organizational structure, but somehow, they also were able to almost single-handedly disrupt the Soviet campaigns.

This supposed omnipotence may have explained the worries from Moscow that repressive measures were not being implemented fully from the outset, or perhaps the regional committees did not have the intestinal fortitude to push the program as far as it should go. A report from the First Secretary of the All-Union Central Council of Trade Unions Shvernik directly to Stalin from his trip to Kazakhstan tried to ascertain why grain procurement was failing, and he attributed it to insufficient use of repression: in fact, Shvernik noted that "the late use of repression… led to kulaks and rich peasants selling some of their grain at the market or hiding it in pits, while they themselves fled to nearby construction sites and state farms. Repressive measures against the kulaks were also sometimes weak. In some village councils of the Kustanai district, kulaks were sentenced to forced labor… giving the kulaks the opportunity to further expand their agitation against grain procurements" (APRF, 1931, March 12). On the other hand, numerous reports seemed to excuse the aforementioned "excesses" of the drives, showing that Moscow's fears were unfounded, as in an OGPU report in 1931 where they noted that "during the grain procurement campaigns of the autumn-winter period of 1930, numerous facts were recorded: beatings of the poor and farm laborers by the authorized representatives of the procurement campaigns, a number of cases of stripping naked in the cold, dousing with cold water, mass arrests of the poor with the detention of those arrested in cold rooms, without food, etc." (CAFSBRF, 1931, January 12). The result of these "excesses" was often the arrest and/or execution of the party official involved, a poor trade of one communist for countless families upended or destroyed.

Indeed, the human toll from these repressions was immense and, on the basis of archival arrest records, it appears that the caricature of the wily nomad was less accurate than the reality of an organizational form under assault from the outside struggling to survive. How else can one explain arrests such as Alikan Belyubaev, a nomad from the East Kazakhstan region, Tarbagatai district, who was arrested on September 11, 1930 by the border guard service, and shot on March 10, 1931 – sometime before his 11th birthday?[7] Or Sarmartian Sultambekov, an illiterate nomad from Shimkent, who was arrested by the OGPU on June 10, 1930, and shot a month later?[8] Arrest records also can show where OGPU detachments were conducting specific repressions, as in the East Kazakhstan region, Chingistau district, aul 7, where Tumek Dumalakov (nomadic pastoralist), Sagidula Esemkanov (also a nomadic pastoralist), and Shaynada Tastaganov (again, a nomadic pastoralist)[9] were all arrested on the same day and sentenced on the same day (June 27, 1931) to various stints in labor camps. Repatriated nomads were also targeted, as with Manen

---

[7] See the Portal of the State Commission for the Complete Rehabilitation of Victims of Political Repression, Republic of Kazakhstan, https://e-memory.kz/en/search/lagerya/?ID=111785/beljubaev-alikan.
[8] https://e-memory.kz/en/search/lagerya/?ID=126584/sultambekov-sarmat.
[9] https://base.memo.ru/person/show/3019009.

Mukhamadiev, another illiterate nomadic cattle herder who had fled to Xinjiang and was then detained attempting to recross the border with his cattle – and sentenced to 10 years in a corrective labor camp.[10]

Perhaps because of the human toll, the nomads kept attempting to wander away and once again the Bolsheviks attempted to restrict their movement the only way they knew how, i.e., with more force. Communist Party documents throughout 1933 remained consumed with preventing any movement of the nomads, keeping them circumscribed to collective farms no matter what the cost. For example, an internal Bolshevik document proclaimed that the true problem was that "the leading [Bolshevik] officials did not assimilate the economic and political significance of *suspending the migrations*, especially in connection with the preparations for the sowing season, continuing their indifferent attitude to the settlement of the nomads, allowing conciliation to be made by the counter-revolutionary, kulak-Bai elements, who used the migrations to agitate against [grain procurement] measures" (RGASPI, 1933, February 12). And of course, after this castigation, border guards and the OGPU could not afford to countenance any type of migration, with the OGPU proudly proclaiming in late 1933 that "in the first ten days of September, 1.5 thousand farms of nomads intending to leave for Afghanistan were detained in the border zone" (CAFSBRF, 1933, October 20).

The obsessive focus on ending migration was constant throughout even the later years of the sedentarization drive, as the famine began to dissipate and hundreds of thousands had died, but another reality was slowly dawning on the Bolsheviks: having brought back hundreds of thousands of nomads from China or Western Siberia and having successfully forced others to remain in place, it was soon discovered that the Kazakh ASSR had little organization or supplies to support them. Put a different way, once the key Soviet objective had been achieved – forcing the nomads to remain where food was scarce (RGASPI, 1932, July 2) – no other thought had really been given as to how the nomads would be made into sedentarized peasants. For example, a memorandum in February 1933 entitled "On the Situation of Nomads" bluntly stated that "With the onset of warmth, mass returns of nomads… is possible, following the example of previous years. The difficulty of the situation is not only in the presence of 100,000 nomads, but also in the fact that at least the same number of farms on the ground are actually starving and need urgent help" (CSARK, 1933, February). A further memorandum the following month stated more explicitly, "due to the fact that neighboring regions do not stop sending the nomads back, we ask the Central Committee to oblige Central Asia and Siberia to organize the settlement of the nomads in the localities and to stop sending them further to Kazakhstan" (APRK, 1933, March 29).

[10] Found on https://ru.openlist.wiki/.

The results of the forced migration to the collective farms in the absence of any infrastructure to support them was predictable: as noted above, nomads suffered disproportionately from disease, which their forced migration onto collective farms worsened, with a top-secret report of the Kazakh Regional Committee noting that "nomads and especially homeless people…make up to 80% of the total number of smallpox cases" (APRK, 1933, June 1). In the East Kazakhstan region, officials saw a cycle of despair, as nomads "find themselves in extremely difficult conditions, from chronic malnutrition among the nomads, mortality on the ground of hunger and diseases is growing, which in turn should cause repeated migrations" (CDNI, 1933, March). In fact, the inability of the Soviet machines to switch from perpetrating evil to actually protecting the sedentarized people can be seen in a threatening letter from the Almaty region executive district to the Leninsky Village Council, which showed what sorts of services nomads could expect when they returned to the Soviet Union:

> *Recently, there have been cases of mortality of nomads walking along the Kaskelen-Alma-Ata highway, and therefore it is proposed to you: upon receipt of this, immediately establish a daily duty of carts along the tract from the city of Alma-Atado, Kaskelen station, alternating with the Kaskelen village council. Your cart on duty must constantly travel along the highway from Kaskelen to Alma-Ata and back during the whole day, and all exhausted or exhausted and poorly dressed people (nomads), i.e. those who cannot walk, are immediately put on the cart and presented to the feeding point in the station of Kaskelen, informing the Kaskelen village council about this. In the case of an accident, you are on duty and if bodies are discovered, you will be held accountable* (CSARK, 1933, January 20).

In other words, after forced migration and closure of normal nomadic routes and routines, at least the Party would be there (hopefully) to pick up your corpse from the side of the road.

**Discussion**

As shown above, the Soviet apparatus was obsessed with controlling the forms of economic institutions within the country, believing that collectivization and sedentarization would increase grain production and demonstrate the superiority of communism. With a single-minded focus on the organizational arrangements in the agricultural sector, the Soviet machine was convinced it could impose its preferred organizational form on nomads throughout Central Asia but especially in Kazakhstan, which they had specific plans for related to food production. When this seemingly simple task failed from the outset, the bureaucratic machines (already primed in this role), rather than changing course, smoothly transitioned into the administrative evil of Balfour et al. (2019). Historians date the end of the sedentarization project concurrently with the end of collectivization, generally around late 1933 and 1934, with the denouement being Goloshchekin's removal as First Secretary in January 1933 (due to the failures noted above) and the end credits rolling with the cessation of the famine in 1934. But even with the machines running at all levels and every single day, the ultimate destiny of the project was failure across any metric. In the first instance, in terms of human cost, it was an avoidable tragedy on a massive scale – based on census data, historians and the Soviets themselves estimated that 1.3 million Kazakhs, a quarter of the entire population of the ASSR, died in the famine, while a presidential commission under Nursultan Nazarbayev in 1992 claimed up to 2.2 million people in total were killed (Cameron, 2023). Rather than sedentarizing the nomads to be productive contributors to the Soviet economy, the nomads were sedentarized through death and detention.

Similarly, by the Soviets' own goals, the results from the twin programs of collectivization and sedentarization also failed miserably. Historians have reassembled grain and livestock production data before, during, and after the drive in Central Asia, and the numbers attributable to Soviet policies are astounding: Pianciola (2001:242) notes that "between 1928 and 1934, Kazakh livestock as a percentage of the Soviet total declined from 18 to 4.5 percent. For the entire USSR between 1928 and 1933, the number of cattle fell by 44 percent, and of sheep and goats by as much as 65 percent; for Kazakhstan, the respective figures were 79 percent and 90 percent." In another compilation of Soviet statistics, Pianciola (2004) quotes archival materials that show that the land under cultivation in Kazakhstan actually fell by 4.2% from 1930 to 1934. Even production of crops that Kazakhstan was suited for, such as hay, fell from 4.42 million tons in 1928 to 2.73 million tons in 1933 (according to raw data from Davies and Wheatcroft [2004]).

The problems did not only exist in the outputs and outcomes but also in the inputs that the Soviets provided. Ironically, in a system built on statistics, reports, and paperwork flying in all directions, even the Soviets had no way to measure how sedentarization was proceeding at the time and at what magnitude. In the words of Pianciola (2001:242), "between 1930 and 1934 regional officials had no clear idea of how many people were actually present in the republic." A commission headed by Aleksei Kiselev, Secretary of the Central Executive Committee of the RSFSR, was convened in 1928 and again in 1934 to understand the problems that kept reoccurring in the Kazakh ASSR specifically with regard to the nomads. In a draft report for the Central Committee written after the famine, Kiselev pointed the finger at Goloshchekin for not knowing (or misleading the Committee) about how sedentarization was proceeding:

> *In 1932 Comrade Goloshchekin [...] furnished the Central Committee with false information, according to which in Kazakstan 230,000 nomad families had been brought under sedentarization, when instead, as the investigation on sedentarization has shown, at the end of 1933 only 70,000 families had been sedentarized* (Quoted in Pianciola [2004:179]).

Even Goloshchekin's made-up numbers were problematic from the point of view of the Bolsheviks, as the plan for the Kazakh ASSR called for sedentarizing 544,000 nomad households by the end of 1933, a far cry from what was achieved and even from what was claimed (Pianciola, 2001). Hartwell (2023), citing Soviet statistics but questioning their validity, notes that the Soviets claimed to have settled 338,685 nomad households by 1937, a massive disruption if true but still only 62% of the plan – and a shadow of the number which were killed and/or arrested to get to that point.

Counting the number of nomads sedentarized was one failure of the drive, but the deeper one was the clash of organizations and in particular the misunderstanding, or rather, of willful ignorance, of the mechanisms underneath nomadism. Like the entire communist experiment, where the Bolsheviks thought they were the "engineers of human souls" (Škvorecký, 1999), the Bolsheviks had a misguided sense of what their power was *vis a vis* the complex system of nomadism; in particular, they believed they could easily, through brute force, stop the nomads from… being nomads. Indeed, a difficulty with the ability of organizations to impose their will on others ironically comes back to the same attribute of a system which may enable power imbalances across organizations and institutions: the complexity of large-scale adaptive systems (Child & Rodrigues, 2011). Organizations may perceive external threats as easily overcome without understanding the underlying intricacies of what is driving these threats, underestimating the complexity of the threat and thus causing their own difficulties in overcoming that threat.

This is precisely what occurred in the Soviet case, as "the nomadic culture of the Kazakh people was perceived by Soviet functionaries as backward due to the difficulty of understanding nomadic society and the importance it attached to ancestral genealogical origin and descent groups or clans" (Sailaubay & Zhanbossinova, 2024:1). Despite being "backward," it continued to adapt to Soviet pressure, and this resilience was even recognized in real time as it was happening. The OGPU noted in March 1933 (CDNI, 1933, March) that the attempts to forcibly migrate the nomads to collective farms just created outbreaks of disease and starvation that precipitated more migration and a reliance on traditional nomadic forms of survival. This then pushed the Soviet organs into their traditional roles, increasing repression as the standard operating procedure of the system; however, this became even more counterproductive to achieving the goals of the program, setting off a spiral that only encouraged nomadism. The only way in which the Soviets were able to overcome the complexity of nomadism was through repression that killed hundreds of thousands of nomads, pressure which "exploded the traditional structures of the Kazakh village" (Yesdauletova et al., 2015:543).

But while the Soviets did succeed in some manner in breaking the backs of resistance of nomads (Hartwell, 2023), and whereas the modern world and the nation-state were already combining to make nomadism an anachronistic form of organization in Kazakhstan, the final word in this saga should not come from our assessment of the devil from the machine at a vantage point of almost a century removed from the policies of 1928 to 1933. Instead, there was a much more proximate admission of failure, as the Soviets themselves never truly vanquished the complexity of nomadism. In fact, they were to quietly reverse many of their sedentarization policies just a few years after the events depicted in this paper. Despite the elimination of over a million Kazakhs during their ill-fated approach to break nomadism, the Soviets finally relented in the face of a much more pressing challenge; on the eve of World War Two, when the Soviets' erstwhile allies in Nazi Germany turned from Western Europe and invaded the USSR, the Central Committee shifted its stance on nomadism and permitted long distance pasture-based nomadism for livestock management (Alimaev & Behnke, 2008). It appeared that nomadism, if done for the survival of the state rather than for the survival of nomads themselves, was an acceptable form of organizing.

In this sense, the machines did have an "off" switch, but it was only thrown when the Party felt that it needed to be.

## Conclusion

This paper has explored what happens when organizations are confronted by perceived disorder in their external environment and how their attempts to impose order can lead to the creation of "the devil from the machine." In particular, we explored this clash in the context of forced migrations, namely the Soviet drive in Kazakhstan in the late 1920s and early 1930s to force nomads onto collective farms and turn them into sedentary farmers. Using archival documents compiled at all levels of the Communist Party, our examination has shown how the single-minded focus on imposing order onto perceived chaos drove the Party farther and farther into relying on its standard operating procedures, unable to understand the complexities of nomadism. Unfortunately, these procedures were overwhelming repression, as all levels of the Party attempted to overcome nomadism as an organization and social structure by brute force. This approach was typified by Stalin's reply to Lady Astor's question in Moscow in 1931, "When are you going to stop killing people?" Stalin stared at her and said softly, "When it is no longer necessary" (Death in the Kremlin, 1953, March 16).

Given the focus of this paper on a historical episode in a country that no longer exists, this examination necessarily has some limitations. In the first instance, generalization to other cases may be difficult, especially given the resources that the Soviet state and the Bolsheviks had at their disposal to attempt to sedentarize the nomads. Indeed, the sheer amount of state power which was brought to bear in this instance was a capacity which many states in emerging markets or less developed countries simply do not have, making this case less applicable for understanding why organizations can choose this path in a disordered field. Similarly, this paper focuses on the Kazakh nomads because of their scale and because of the interest that the Soviet machinery had in them in particular. However, Kazakhs were not the only nomads in Central Asia, and the Soviet choice to sedentarize the Kazakhs was based on deeper plans for the region to supply Moscow; in other areas of Central Asia, such as Uzbekistan, cotton farming was chosen for the native population, leading to different approaches to the nomadic population (where far less was written and/or documented). Thus, our story only picks up somewhere in the middle, after the economic decisions had already been made, and comparing the Kazakhs with other Central Asian nomads may have illuminated how the machines of forced migration worked in a more complete way.

Nevertheless, while there are limitations to this work, I believe this case also opens up fascinating possibilities for researchers in organizational theory, both related to this specific case and beyond. With reference to the Soviet case, Soviet archives, despite increasing inaccessibility at key centers in Russia, remain a treasure trove of understanding the machines of evil, and the ongoing digitalization of arrest records being carried out in Kazakhstan will help us to better understand the human victims of the famine of 1930-1933. The extensive archival materials consulted here are just the tip of the iceberg and require the same scrutiny as has been given the genocidal Ukrainian famine during the same years (Cameron, 2018), with perhaps a greater emphasis on arrest records than I have done here. In this way, we will be able to see the human cost from the side of the nomads rather than just the oppressors. And, as just noted, researchers can also begin to look into Uzbkeistan, Kyrgyzstan, and Turkmenistan to see if their experiences with nomads during the Soviet Union are similar to or have crucial differences with the Kazakh case.

Moreover, the Soviet case is not an isolated one in terms of restricting nomadism, as evidenced by recent scholarship, and moving beyond the Soviet experience to more contemporary cases of forced migration of nomads will also help to understand how political organizations attempt to make sense of disorder when they do not have the same resources or power that the Bolsheviks had. How do organizations closer in power levels to nomadism accommodate, resist, or pressure nomadism to adapt – and are they successful? And are there complexities in public administration that also come to bear in this interplay of organizations? Indeed, in a world full of migration, migration reversals, forced migration, and geopolitical tension, the question of what drives the "devil from the machine" appears to be of increasing relevance and can keep researchers occupied for years.

ARCHIVE ABBREVIATIONS

| | |
|---|---|
| APRF | Archive of the President of the Russian Federation |
| APRK | Archive of the President of the Republic of Kazakhstan |
| CAFSBRF | Central Archive of the Federal Security Bureau of the Russian Federation |
| CDNI | Center for Documentation of Contemporary History of the East Kazakhstan Region |
| CSARK | Central State Archive of the Republic of Kazakhstan |
| GARF | Central State Archive of the Russian Federation |
| RGASPI | Russian State Archive of Social and Political History |
| SAWKR | State Archive of the West Kazakhstan Region |

ARCHIVAL DOCUMENTS AND COLLECTIONS

**Figure 1 – Report of the Middle Volga Regional Executive Committee on the mass arrival of Kazakhs in the Middle Volga region. February 12, 1933**

Копия 118

Р.С.Ф.С.Р.
ИСПОЛНИТЕЛЬНЫЙ КОМИТЕТ
С О В Е Т О В
Рабочих, Крестьянских,
Красноармейских и Казачьих
Д е п у т а т о в
СРЕДНЕ-ВОЛЖСКОГО
к р а я.
12 февраля 1933 года
№ 33/9.
гор. САМАРА.

В СОВЕТ НАРОДНЫХ КОМИССАРОВ
РСФСР.
гор. МОСКВА.

Начиная с 1930-31 года из пределов Казакстана на территорию Средней Волги начался большой наплыв казаков и, этот наплыв достиг громадных размеров весною 1932 года. Всего казаков в нашем крае тогда насчитывалось свыше 60 тыс. человек.

Положение казаков было чрезвычайно тяжелое. Они загружали вокзалы Соль-Илецка, Оренбурга и других станций. Среди казаков развивалось эпидемическое заболевание: сыпной тиф, брюшной тиф, оспа и т.п.

Крайисполкомом и непосредственно Райисполкомами было принято ряд конкретных мероприятий в деле улучшения положения казаков. Специально проводилась работа по вербовке их в качестве рабочих в совхозы, промпредприятия и новостройки; были размещены почти все казаки по районам; ликвидирована беспризорность среди детей; оказана должная как денежная, та[к] и продовольственная помощь.

Кроме того, в связи с принятым решением центральных органов о реэвакуации казаков бы